\documentclass[10pt,conference]{IEEEtran}
\usepackage{latexsym}
\usepackage{epsfig}
\usepackage{times}
\usepackage{amsfonts} 
\usepackage{float}
\usepackage{multicol}
\begin{document}


\newtheorem{teor}{Theorem}[section]
\newtheorem{examp}[teor]{Example}
\newtheorem{con}[teor]{Conjecture}
\newtheorem{prop}[teor]{Proposition}
\newtheorem{theo}[teor]{Theorem}
\newtheorem{cor}[teor]{Corollary}
\newtheorem{defi}[teor]{Definition}
\newtheorem{lem}[teor]{Lemma}
\newtheorem{rem}[teor]{Remark}
\newcommand{\beq}{\begin{equation}}
\newcommand{\eeq}{\end{equation}}
\newcommand{\beql}[1]{\begin{equation} \label{#1}}
\newcommand{\eeql}{\end{equation}}
\newcommand{\beqa}{\begin{eqnarray*}}
\newcommand{\eeqa}{\end{eqnarray*}}
\newcommand{\beqal}[1]{\begin{eqnarray} \label{#1}}
\newcommand{\eeqal}{\end{eqnarray}}
\newcommand{\beqan}{\begin{eqnarray}}
\newcommand{\eeqan}{\end{eqnarray}}
\newcommand{\bpf}{\begin{proof}}
\newcommand{\epf}{\end{proof}}
\newcommand{\bF}{{\bf F}}
\newcommand{\cA}{{\cal A}}
\newcommand{\cB}{{\cal B}}
\newcommand{\cC}{{\cal C}}
\newcommand{\cD}{{\cal D}}
\newcommand{\cH}{{\cal H}}
\newcommand{\cM}{{\cal M}}
\newcommand{\supp}{{\rm supp}}
\newcommand{\bfh}{{\bf h}}
\newcommand{\bfx}{{\bf x}}
\newcommand{\bfy}{{\bf y}}
\newcommand{\bfzero}{{\bf 0}}
\newcommand{\bfone}{{\bf 1}}
\newcommand{\wt}{{\rm wt}}
\newcommand{\Prob}{{\rm Prob}}
\newcommand{\bbF}{{\mathbb{F}}}
\newcommand{\spn}{{\rm span}}
\newcommand{\gd}{{\delta}}
\newcommand{\gre}{{\epsilon}}

\title{Generating parity check equations for bounded-distance iterative erasure decoding}
\author{\authorblockN{Henk D.L.\ Hollmann}
\authorblockA{Philips Research Laboratories \\
Prof.\ Holstlaan 4, 5656 AA Eindhoven, The Netherlands \\
Email: henk.d.l.hollmann@philips.com}
\and
\authorblockN{Ludo M.G.M.\ Tolhuizen}
\authorblockA{Philips Research Laboratories \\
Prof.\ Holstlaan 4, 5656 AA Eindhoven, The Netherlands
\\ Email: ludo.tolhuizen@philips.com}}
%

\maketitle

\begin{abstract}
\noindent A generic $(r,m)$-erasure correcting set is a collection
of vectors in $\bF_2^r$ which can be used to generate, for each
binary linear code of codimension $r$, a collection of parity
check equations that enables iterative decoding of all correctable
erasure patterns of size at most $m$. That is to say, the only
stopping sets of size at most $m$ for the
generated parity check equations are  the
erasure patterns for which there is more than one manner to fill in the
erasures to obtain a codeword.

We give an explicit construction of generic $(r,m)$-erasure
correcting sets of cardinality $\sum_{i=0}^{m-1} {r-1\choose i}$.
Using a random-coding-like argument, we show that for fixed $m$,
the minimum size of a generic $(r,m)$-erasure correcting set is
linear
in $r$. \\ \\
{\bf Keywords}: iterative decoding, binary erasure channel,
stopping set
\end{abstract}



%

%

%

%

\section{Introduction\label{Sec:int}}
This paper is motivated by the following well-known scheme for
iterative decoding of a binary linear code $C$ used on the binary
erasure channel \cite{Di02}. We are given a set ${\cal H}$ of
parity check equations for $C$. For a received word with $E$ as
set of erased positions, we inspect if one of the parity check
equations from ${\cal H}$ involves exactly one of the erasures. If
so, we determine the value of the erasure involved in this
equation and continue; if not, we stop the algorithm. In the
latter case, the set $E$ is called a {\em stopping set for ${\cal
H}$} \cite{Di02}. Different sets ${\cal H}$ of parity check
equations for $C$ may result in different stopping sets. Note,
however, that the support of a codeword is always a stopping set,
as by definition each parity check vector has an even number of
ones within the support of a codeword.

We are interested in the behavior of the iterative decoding
algorithm for erasure patterns that are {\em $C$-correctable},
i.e., for which there is only one way to fill in the erasures to
obtain a word from $C$. In fact, we wish to find parity check
collections with which the iterative decoding algorithm decodes
all $C$-correctable patterns of a sufficiently small cardinality.
As related work, we mention that in \cite{WeAb05a}, Weber and
Abdel-Ghaffar construct collections of parity check equations for
the Hamming code $C_r$ of redundancy $r$ with which the iterative
algorithm decodes all $C_r$-correctable erasure patterns of size
at most 3. In \cite{ScVa05} and \cite{ScVa06}, Schwartz and Vardy
study the minimum size of collections of parity check equations
for a code $C$ for which the iterative algorithm decodes all
erasure patterns of size less than the minimum distance of $C$
(note that all such erasure patterns are $C$-correctable).

 In \cite{ht1}, we introduced
and constructed so-called {\em generic (r,m)-erasure correcting
sets}. These are subsets ${\cal A}$ of $\mathbb{F}_2^r$ such that
for {\em any} code $C$ of length $n$ and codimension $r$, and {\em
any} $r\times n$ parity check matrix $H$ for this code, the
collection of parity check equations
\[ \{ {\bf a}H \mid {\bf a}\in {\cal A}\} \]
allows the iterative decoding algorithm to correct all
$C$-correctable erasure patterns of size at most $m$. Our aim is
to construct generic
$(r,m)$-erasure correcting sets of small size, and to investigate
the minimum size $F(r,m)$ of such sets.
At first sight, the definition of generic $(r,m)$-erasure
correcting sets seems to be very restrictive. However, in
\cite{ht1} it was shown that if a set of linear combinations works
for a parity check matrix $H_r$ of the Hamming code of redundancy
$r$, then it works for {\em any\/} parity matrix for any code of
redundancy $r$ -- see Proposition~2.6 of the present paper for a
more precise formulation and a proof.

The paper is organized as follows. In Section~\ref{Sec:pre}, we
introduce notations and definitions. In Section~\ref{Sec:constr},
we present the explicit generic $(r,m)$-erasure correcting sets
from \cite{ht1}. These sets have size $\sum_{i=0}^{m-1} {r-1
\choose i}$.
 In Section~\ref{Sec:prob} we  show that $F(r,m)\geq r$. With a random-coding like
 argument we show that for each $m\geq 1$, there exist a constant $c_m$ such that
$F(r,m)\leq c_mr$ for each $r\geq m$. At present we do not have a
{\em constructive\/} proof that $F(r,m)$ is linear in $r$.
\footnote{In \cite{ht2}, we gave an explicit recursive
construction, too involved to be included here, of generic
$(r,3)$-erasure correcting sets of size 1+3($r-1$)$^{\log_2 3}$.}

\section{Preliminaries \label{Sec:pre}}
In this section, we introduce some notations and definitions.
Throughout this paper, we use boldface letters to denote row
vectors. All vectors and matrices are binary. If there is no
confusion about the length of vectors, we denote with ${\bf 0}$
the vector consisting of only zeroes,
and with ${\bf e}_i$ the $i$-th unit vector, the vector that has a
one in position $i$ and zeroes elsewhere.

The size of a set $A$ is  denoted by $|A|$.  If $H$ is a $r\times
n$ matrix and $E\subseteq\{1,2,\ldots ,n\}$, then the {\em
restriction\/ $H(E)$ of $H$ to $E$\/} denotes the $r\times |E|$
matrix consisting of those columns of $H$ indexed by $E$.
Similarly, if ${\bf x}\in\mathbb{F}_2^n$ and
$E\subseteq\{1,2,\ldots ,n\}$, then the restriction ${\bf x}(E)$
of ${\bf x}$ to $E$ is the vector of length $|E|$ consisting of
the entries indexed by $E$.

The {\em support\/} supp({\bf x}) of a vector ${\bf x}\in \mathbb{F}_2^n$
is the set of its non-zero coordinates, that is,
\[ \mbox{supp}({\bf x}) = \{ i\in \{1,2,\ldots ,n\} \mid x_i\neq 0\} , \]
and the weight wt({\bf x}) of {\bf x} is the size $|\supp(\bfx)|$ of its support.

As usual, an $[n,k]$ code $C$ is a $k$-dimensional subspace of
$\mathbb{F}_2^n$; the dual code of $C$, denoted by $C^{\perp}$, is
the $[n,r]$ code with $r=n-k$ consisting of all vectors in $\mathbb{F}_2^n$
that have inner product 0 with all words from $C$. The number $r$ is referred to as
the {\em codimension\/} or {\em redundancy\/} of the code. An $r\times n$
matrix is called a parity check matrix for $C$ if its rows span
$C^{\perp}$. When we speak about ``code'', we will always mean binary linear code.

The following definitions are taken from \cite{ht1}. \vspace*{2mm}
\begin{defi}
Let $C\subseteq \bF_2^n$ be a code. A set $E\subseteq \{1,2,\ldots
,n\}$ is called {\em $C$-uncorrectable\/} if it contains the
support of a non-zero codeword, and {\em $C$-correctable\/}
otherwise.
\end{defi}
\vspace*{2mm}
 The motivation for this definition is that a
received word containing only correct symbols and erasures can be
decoded unambiguously precisely when exactly one codeword agrees
with this word in the non-erased positions; for linear codes this
is the case precisely when the set of erasures does not contain
the support of a non-zero codeword.
\vspace*{2mm}
\begin{defi}\label{def22}
Let $C\subseteq\mathbb{F}_2^n$ be a code. A set ${\cal H}\subseteq
C^{\perp}$ is called m-{\em erasure reducing for $C$} if for each
erasure pattern $E\subseteq\{1,2,\ldots ,n\}$ of size $m$ that is
$C$-correctable, there exists a parity check equation ${\bf
h}$$\in$${\cal H}$ with exactly one 1 in the positions indexed by
$E$, that is, with wt$({\bf h}(E))$=1. The set ${\cal H}$ is
called m-{\em erasure decoding for $C$} if it is $m'$-erasure
reducing for $C$ for all $m'$ with 1$\leq m'\leq m$.
\end{defi}
\vspace*{2mm} Definition~\ref{def22} implies the following. If the
iterative decoding algorithm is used with a set ${\cal H}$ of
parity check equations that is $m$-erasure reducing for $C$, then
for each $C$-correctable erasure pattern of size $m$ at least one
erasure is resolved; if ${\cal H}$ is $m$-erasure correcting for
$C$, then the iterative decoding algorithm can correct all
$C$-correctable erasure patterns of size at most $m$ by removing
one erasure at the time, without ever getting stuck.
Note that this definition makes no requirements on the behaviour of the
iterative
decoding algorithm for erasure patterns that are {\em not\/} $C$-correctable.

The following example shows that an $m$-erasure reducing set for a
code $C$ need not be an $m$-erasure correcting set for $C$.
\vspace*{2mm}
\begin{examp}\label{Ex1}
Let $C$ be the binary $[$5,1,5$]$ repetition code,
      and let
      ${\cal H}$ consist of the four vectors ${\bf h}_1=10001$, {\bf h}$_2=01100$,
      ${\bf h}_3=01111$, and ${\bf h}_4=01010$. Note that $\cH$ spans the dual code
$C^\perp$ of $C$ (which is just the even-weight code of length
five). In the table
      below, we provide for each set of erasures of size four a parity check
      equation that has weight one inside this erasure set.
      \vspace*{2mm}

      \begin{tabular}{cc}
      non-erased position & parity check equation  \\
      1   & ${\bf h}_1$    \\
      2   & ${\bf h}_2$   \\
      3   & ${\bf h}_2$   \\
      4   & ${\bf h}_4$   \\
      5   & ${\bf h}_1$
      \end{tabular}

      The set ${\cal H}$ is therefore 4-erasure reducing for $C$.
      It is, however, not 4-erasure correcting for $C$, as
      $\{2,3,4\}$ is a stopping set that does not
      contain the support of a nonzero codeword.
So for example the erasure set $\{1,2,3,4\}$ is $C$-correctable,
and can be reduced but not corrected by $\cH$. \vspace*{2mm}
\end{examp}

 Finally, in \cite{ht1} we introduced the notion of a ``generic''
$m$-erasure correcting and reducing set for codes of a fixed
codimension. The idea is to describe which linear combinations to
take given any parity check matrix for any such code.
\vspace*{2mm}
\begin{defi}
   Let $1\leq m\leq r$. A set ${\cal A}\subseteq \mathbb{F}_2^r$ is
   called {\em generic $(r,m)$-erasure reducing\/} if for any
   $n\geq r$ and for any $r\times n$  matrix $H$ of rank~$r$,
    the collection
    $\{{\bf a}H \mid {\bf a}\in {\cal A}\}$
     is $m$-erasure reducing for the code with parity check matrix
     $H$. \\
 The set ${\cal A}\subseteq \mathbb{F}_2^r$ is
   called {\em generic $(r,m)$-erasure correcting } if
it is generic $(r,m')$-erasure reducing for all $m'$ with $1\leq m'\leq m$.
     \end{defi}
\vspace*{2mm}The following useful characterization of generic
$(r,m)$-erasure reducing sets has been obtained in \cite{ht1}.
\vspace*{2mm}
 \begin{prop}\label{altdef}
   A set ${\cal A}\subseteq\mathbb{F}_2^r$ is generic $(r,m)$-erasure reducing if
and only if for any $r\times m$ matrix $M$ of rank
   $m$ there is a vector {\bf a}$\in$${\cal A}$ such that
   wt({\bf a}$M)=1$.
   \end{prop}
   \vspace*{2mm}
The proof of this proposition can be outlined as follows.  It can
be shown that an erasure pattern $E$ is $C$-correctable if and
only if for any parity check matrix $H$ for $C$, the restriction
$H(E)$ has full rank. Hence, we need only consider $r\times m$
submatrices of full rank, and each $r\times m$ matrix of full rank
can occur as such a submatrix.
Using Propostion~\ref{altdef}, it  can be shown that for all codes
of a fixed codimension, the Hamming code is the most difficult
code to design generic erasure reducing sets for. The following
proposition states this fact more precisely. \vspace*{2mm}
\begin{prop} Let $m\leq r$. Let $H_r$ be a parity check matrix of
the $[2^r-1,2^r-r-1,3]$ Hamming code $C_r$. A set ${\cal
A}\subset\mathbb{F}_2^r$ is generic $(r,m)$-erasure reducing if
and only if ${\cal H}=\{ {\bf a}H_r\mid {\bf a}\in {\cal A}\}$ is
$m$-erasure reducing for $C_r$.
\end{prop}
\bpf Combination of Proposition~\ref{altdef} and the fact that any
$r\times m$ matrix of rank $m$ occurs, up to a column permutation,
as a submatrix of $H_r$. \vspace*{2mm} \epf
 We are interested in
generic $(r,m)$-erasure reducing sets of small size. This
motivates the following definition. \vspace*{2mm}
\begin{defi}
For $1\leq m\leq r$, we define $F(r,m)$ as the smallest size of
any generic $(r,m)$-erasure reducing set.
\end{defi}
\vspace*{2mm}
 Note that Proposition~\ref{altdef} implies that
$\mathbb{F}_2^r\setminus\{ {\bf 0}\}$ is generic $(r,m)$-erasure
reducing, so $F(r,m)$ is well-defined.
%
\vspace*{1mm} \\
 Example~\ref{Ex1} shows that for particular codes $C$, the notions "$m$-erasure reducing for C"
 and "$m$-erasure correcting for $C$" need not be the same; as we proceed to show, the
notions ``generic $(r,m)$-erasure reducing'' and ``generic
$(r,m)$-erasure correcting'', are, somewhat surprisingly,
equivalent. \vspace*{2mm}

\begin{prop}\label{mthenm-1}
Let $2\leq m\leq r$. A generic $(r,m)$-erasure reducing set is a
generic $(r,m-1)$-erasure reducing set. \vspace*{2mm}
\end{prop}
\bpf Let ${\cal A}$ be a generic $(r,m)$-erasure-reducing set.
Let $M$ be a binary  $r\times (m-1)$ matrix
of rank $m-1$. We write \[ M= \left[ M_0 \mid {\bf x}^\top \right],
\] where ${\bf x}^\top$ denotes the rightmost column of $M$. Let
${\bf y}^\top$ be a vector in $\mathbb{F}_2^r$ that is not in the
linear span of the columns of $M$, and let $M^{\prime}$ denote the
$r\times m$ matrix defined as
\[ M^{\prime} = \left[ M_0 \mid {\bf y}^\top  \mid {\bf x}^\top+{\bf y}^\top \right] . \]
As $M^{\prime}$ has rank $m$, there exists a vector {\bf
a}$\in$${\cal A}$ such that wt$({\bf a}M^{\prime})=1$. We claim
that wt$({\bf a}M)=1$. This is clear if wt$({\bf a}M_0)=1$, as
then ${\bf a}{\bf x}^\top={\bf a}{\bf y}^\top=0$. If ${\bf
a}M_0=0$, then {\bf a}{\bf y}$^\top=0$ and ${\bf a}({\bf
x}^\top+{\bf y}^\top)=1$, or vice versa. In either case, {\bf
a}${\bf x}^\top={\bf a}{\bf y}^\top+{\bf a}({\bf x}^\top+{\bf
y}^\top) = 1$, from which we conclude that in this case also ${\bf
a}M$ has weight 1. \vspace*{2mm} \epf Note that
Proposition~\ref{mthenm-1} implies that the parity check equations
induced by a generic $(r,m)$-erasure reducing set can also be used
to resolve an erasure from a correctable erasure set of size
$m-1,m-2,\ldots$ (as shown in Example~\ref{Ex1}, this need not
hold for a specific $m$-erasure reducing set for a specific code).
In other words, the following proposition holds. \vspace*{2mm}
\begin{prop}\label{red=dec}
Any generic $(r,m)$-erasure reducing set is a generic $(r,m)$-erasure correcting set.
\end{prop}
\vspace*{2mm}
According to Proposition~\ref{red=dec}, the terms ``generic
$(r,m)$-erasure reducing'' and ``generic $(r,m)$-erasure
correcting'' can
 be used interchangably. In the sequel, we
use ``correcting'', and base our results on the characterization
given in Proposition~\ref{altdef}.

\section{Explicit generic $(r,m)$-erasure correcting sets}\label{Sec:constr}
%
 In this section, we describe generic $(r,m)$-erasure correcting sets
${\cal A}_{r,m}$ for all $r$ and $m$ with $r\geq m\geq 2$ (see
also \cite{ht1}).  We will show that the sets ${\cal A}_{r,3}$ are
closely related to the sets found by Weber and Abdel-Ghaffar
\cite{WeAb05a}. Modifications and generalizations of these sets
can be found in \cite{ht1} and \cite{ht2}. \vspace*{2mm}
\begin{theo}
Let $2\leq m\leq r$. The set ${\cal A}_{r,m}$ defined as
\[ {\cal A}_{r,m} = \{ {\bf a}= (a_1,a_2,\ldots ,a_r) \in \mathbb{F}_2^r
\mid a_1=1 \mbox{ and wt}({\bf a})\leq m \}\] is a generic
$(r,m)$-erasure correcting set of size
\[ \sum_{i=0}^{m-1} {r-1 \choose i} . \]
\end{theo} \vspace*{2mm}
 \bpf As ${\cal A}_{r,m}$
consists of all vectors that start with a one and have weight at
most $m-1$ in the positions 2,3,\ldots ,$r$, the statement on the
size of ${\cal A}_{r,m}$ is obvious.

In order to show that ${\cal A}_{r,m}$ is indeed generic
$(r,m)$-erasure correcting, we will use Proposition~\ref{altdef}.
So let $M$ be an $r\times m$ matrix of rank $m$. We have to show
that there is a vector {\bf a}$\in$${\cal A}_{r,m}$ such that
wt({\bf a}$M$)=1. To this end, we proceed as follows.
For $1\leq i\leq r$, let {\bf m}$_i$ denote the $i$-th row of $M$.
Let $I\subseteq\{1,2,\ldots ,r\}$ be such that $\{{\bf m}_i \mid
i\in I\}$ forms a basis for $\mathbb{F}_2^m$. We distinguish two
cases.

(i): ${\bf m}_1\neq {\bf 0}$.

\noindent We can and do choose $I$ such that 1$\in I$. The set $\{
\sum_{i\in I} x_i{\bf m}_i\mid ({x_i})_{i\in I}, x_1=0\}$ is an
$(m-1)$-dimensional space and hence cannot contain all unit
vectors. So there exists a vector ${\bf x}=(x_i)_{i\in I}$ with
$x_1=1$ and wt($\sum_{i\in I} x_i{\bf m}_i)=1$. Now, let ${\bf
a}$$\in$$\mathbb{F}_2^r$ be the vector that agrees with ${\bf x}$
in the positions indexed by $I$ and has zeroes elsewhere. Then
$a_1=x_1=1$ and wt({\bf a})$=$wt$({\bf x})\leq m$, hence {\bf
a}$\in$${\cal A}_{r,m}$ and ${\bf a}M=\sum_{i=1}^r a_i{\bf m}_i=
\sum_{i\in I}x_i{\bf m}_i$, so wt$({\bf a}M)=1$.

(ii): ${\bf m}_1 = {\bf 0}$.

\noindent In this case $1\notin I$. As $ \{{\bf m}_i\mid i\in I\}$
forms a basis,
there are independent vectors ${\bf x}(j)=(x_i(j)\mid i\in I)$
such that ${\bf e}_j= \sum_{i\in I} x_i(j){\bf m}_i$ for all $j$.
As there is just one vector ${\bf x}$ of weight $m$, and there are
$m\geq 2$ unit vectors, there is an index $j$ such that wt$({\bf
x}(j))\leq m-1$. Now, let {\bf a} be the vector that agrees with
${\bf x}(j)$ in the positions indexed by $I$, has a ``1'' in the
first position, and zeroes elsewhere. As wt$({\bf x}(j))\leq m-1$,
the vector {\bf a} is in ${\cal A}_{r,m}$. Moreover, we have that
${\bf a}M = \sum_{i=1}^n a_i{\bf m}_i = a_1{\bf m}_1 + \sum_{i\in
I}a_i{\bf m}_i= {\bf 0} + {\bf e}_j={\bf e}_j$. \vspace*{2mm} \epf
We now relate
our result for $m=3$ to that of Weber and Abdel-Ghaffar
\cite{WeAb05a}, which in our terminology states that
\[ {\cal W}_r = \{ {\bf e}_i \mid 1\leq i\leq r\} \cup
                  \{ {\bf e}_1 + {\bf e}_i + {\bf e}_j\mid
                     2\leq i < j \leq r\} \]
                     is generic $(r,3)$-erasure correcting.
To this end, let $S$ be the matrix with the all-one vector as
leftmost column, and with ${\bf e}_j^\top$ as $j$-th column for
$2\leq j\leq r$. Clearly,  $S$ is invertible, and for 2$\leq i\leq
r$ and $2\leq j < k\leq r$, we have that
\[ {\bf e}_1S={\bf e}_1, \qquad
\mbox{$({\bf e}_1+{\bf e}_i)S= {\bf e}_i$}, \mbox{ and } \]
\[ \mbox{$({\bf e}_1+{\bf e}_j+{\bf e}_k)S = {\bf e}_1+{\bf e}_i+{\bf
e}_j$} \] As a consequence, we have that
\[ {\cal W}_r = \{ {\bf a}S \mid {\bf a}\in {\cal A}_{r,3}\} .
\]
So ${\cal W}_r$ and ${\cal A}_{r,3}$ are related via an
element-wise  multiplication with the invertible matrix~$S$.
\section{Upper and lower bounds on $F(r,m)$}\label{Sec:prob}
In this section, we show that $F(r,m)$ is {\em of linear order\/}
in $r$. To be more precise, we will show that for each $m\geq 1$,
there exists a constant $c_m$ such that for each
$r\geq m$, we have that $r\leq F(r,m)\leq c_mr$. \\
Concerning the lower bound, we have the following lemma.
\vspace*{2mm}
\begin{lem}
Any $(r,m)$-erasure decoding set spans $\mathbb{F}_2^r$. As a
consequence, $F(r,m)\geq r$.
\end{lem}
\vspace*{2mm}
\begin{proof}
(cf. \cite{ht1}) Suppose ${\cal A}\subset \mathbb{F}_2^r$ is such
that span$({\cal A})\neq \mathbb{F}_2^r$. We will show that ${\cal
A}$ is not generic $(r,m)$-erasure decoding by constructing an
$r\times m$ matrix $M$ with rank $m$ such that for each ${\bf
a}$$\in$${\cal A}$, the vector ${\bf a}M$ does not
have weight 1 (cf.\ Propostion~\ref{altdef}). \\
Let {\bf v} be a non-zero vector that has inner product 0 with all
words from ${\cal A}$. Let $M$ be an invertible matrix such that
the $i$-th row of $M$ has odd weight if and only if
$i\in\mbox{supp}({\bf v})$, and let ${\bf a}$$\in$${\cal A}$. As
$({\bf v},{\bf a}) = 0$, the vector ${\bf a}M$ is the sum of an
even number of (odd weight) rows of $M$ indexed by integers from
supp({\bf v}), and some (even weight) rows of $M$ indexed by
integers outside supp({\bf v}). As a consequence, ${\bf a}M$ has
even weight. \vspace*{2mm}
\end{proof}
 The proof for the upper bound (cf. \cite{ht2}) can be
 considered to be a
 random-coding argument: we will show that the collection
of all subsets of $\mathbb{F}_2^r$ of a sufficiently large size
contains at least one generic $(r,m)$-erasure correcting set. The
precise result is as follows.
\begin{teor}\label{T-lin}
For all $m\geq1$ and $r\geq m$, we have that
\[ F(r,m)\leq
 \frac{m}{-\log_2(1-m2^{-m})} \cdot r, \] where $\log_2$
denotes the base-2 logarithm.
\end{teor}
\begin{proof}
Let $1\leq m\leq r$. We
write $\cM_{m,r}$ to denote the collection of all binary $r\times
m$ matrices of rank $m$.
Let $N$ be some positive integer. Consider the following
experiment.
We randomly construct a binary $N\times r $ matrix $A$ by setting
each individual entry to zero or to one, each with probability
$1/2$. We interprete this matrix as a sequence of $N$ row vectors
${\bf a}_1, \ldots, {\bf a}_N$, each of length $r$. For each
matrix $M$ in $\cM_{m,r}$, we define the random variable $X_M$ by
\[
X_M = \left\{
        \begin{array}{cc}
        0, & \mbox{if there is an $i $ such that $\wt({\bf a}_iM)=1$;}\\
        1, & \mbox{otherwise.}
        \end{array}
    \right.
\]
So $X_M=0$ if the matrix $M$ is ``good'' with respect to the
vectors ${\bf a}_1, \ldots, {\bf a}_N$, and $X_M=1$ if $M$ is
``bad''. \\ Furthermore, let the random variable $X$ be defined as
\[ X=\sum_{M\in\cM_{m,r}} X_M. \]
The random variable $X$ thus counts the number of bad matrices
with respect to $A$; if $X<1$, then  all matrices are ``good''
with respect to $A$,  so that the collection $\cA=\{{\bf a}_1,
\ldots, {\bf a}_N\}\subseteq \bF_2^r$ satisfies the criterion in
Proposition~\ref{altdef}. Consequently, if $E[X]<1$, then all
matrices in
 ${\cal M}_{m,r}$ are
good with respect to some matrix $A$, and so $F(r,m) \leq N$.

In order to compute $E[X]$, we start by fixing a matrix
$M\in\cM_{m,r}$ and
compute the probability $\Prob(X_M=1)$ that $X_M$ is equal to 1.
As $M$ has full rank, there are, for each $i=1,2,\ldots ,m$,
exactly $2^{r-m}$ vectors ${\bf a}$ such that ${\bf a}M={\bf
e}_i$.
 We conclude that there are $m2^{r-m}$ ``good'' vectors for $M$,
and hence $2^m(1-m2^{-m})$ ``bad'' vectors. Now the matrix $M$ is
bad if all the vectors ${\bf a}_1, \ldots, {\bf a}_N$ are bad; we
conclude that
\[ \Prob(X_M=1) = (1-m2^{-m})^N.  \vspace*{2cm} \]


Since expectation is a linear operation, we have that
\[ E[X]=\sum_{M\in\cM_{m,r}} E[X_M] = |\cM_{m,r}|(1-m2^{-m})^N,\]
from which we conclude that $E[X]<1$ if and only if \beql{E_prob}
N
> \frac{\log_2 |\cM_{m,r}|}{ -\log_2 (1-m2^{-m})}. \eeql

As a consequence of the foregoing, we have that $F(r,m)\leq N$ if
$N$ satisfies (\ref{E_prob}); as $|{\cal M}_{m,r}|$ is at most
2$^{mr}$, the cardinality of the set of all $r\times m$ matrices,
it follows  that $F(m,r)\leq N$ whenever
\[  N \geq \frac{m}{ -\log_2 (1-m2^{-m})} \cdot r . \]
\end{proof}
\section{Conclusions}
In this paper, we have introduced the notion of generic
$(r,m)$-erasure correcting sets in $\mathbb{F}_2^r$; such sets
provide for each binary code $C$ with redundancy $r$ a collection
of parity check equations for $C$ that can be used to iteratively
correct all $C$-correctable erasure patterns of size at most $m$.
We provided an explicit construction of generic $(r,m)$-erasure
correcting sets of size $\sum_{i=0}^{m-1}{r-1\choose i}$,
generalizing the result for $m=3$ from \cite{WeAb05a}. We also
showed, by a random-coding-like argument, that for each fixed $m$,
the minimal size of generic $(r,m)$-erasure correcting sets is
{\em linear in r}.

The main remaining problem is to find {\em explicit} constructions
for $(r,m)$-erasure correcting sets of size linear in $r$,
especially in the first open case $m=3$. In \cite{ht2}, we provide
an explicit recursive construction of $(r,3)$-erasure correcting
sets of cardinality $1+3(r-1)^{\log_2 3}$ -- not linear in $r$,
but much smaller than the $(r,3)$-erasure correcting sets from
Section~III for which the cardinality is quadratic in $r$.

\end{document}